\documentclass[]{article}  

\usepackage{graphicx} 
\usepackage[english]{babel}
\usepackage{amsmath, amssymb, amsfonts,  bm}
\usepackage{mathrsfs}
\usepackage[mathcal]{euscript}
\usepackage{enumerate}
\newtheorem{proposition}{Proposition}
\usepackage{float}
\usepackage{color,graphicx,epstopdf}
\usepackage{dsfont}
\usepackage{hyperref}
\usepackage{cite} 
\usepackage{algorithm} 
\usepackage{algorithmic} 
\usepackage{tikz-cd}
\usepackage{mathtools}
\usepackage{indentfirst} 
\usepackage{subcaption} 
\usepackage{amsmath}
\usepackage{setspace}
\newtheorem{theorem}{Theorem}
\newtheorem{definition}{Definition}
\newtheorem{remark}{Remark}
\newtheorem{assumption}{Assumption}

\DeclareFontFamily{OT1}{pzc}{}
\DeclareFontShape{OT1}{pzc}{m}{it}{<-> s * [1.200] pzcmi7t}{}
\DeclareMathAlphabet{\mathpzc}{OT1}{pzc}{m}{it}

\newcommand*\mcapinn[2]{\vcenter{\hbox{$\mathsurround=0pt
  \ifx\displaystyle#1\textstyle\else#1\fi\bigcap$}}}

\newcommand*\mcupinn[2]{\vcenter{\hbox{$\mathsurround=0pt
  \ifx\displaystyle#1\textstyle\else#1\fi\bigcup$}}}

\def\begequarr{\begin{eqnarray}}
\def\endequarr{\end{eqnarray}}
\def\begequarrs{\begin{eqnarray*}}
\def\endequarrs{\end{eqnarray*}}
\def\begequ{\begin{equation}}
\def\endequ{\end{equation}}
\def\begequs{\begin{equation*}}
\def\endequs{\end{equation*}}
\def\begite{\begin{itemize}}
\def\endite{\end{itemize}}

\def\begcen{\begin{center}}
\def\endcen{\end{center}}
\def\begrem{\begin{remark}\rm}
\def\endrem{\end{remark}}
\def\ba{\begin{array}}
\def\ea{\end{array}}

\newcommand{\mV}{\mathrm{V}}
\newcommand{\mG}{\mathrm{G}}
\newcommand{\mE}{\mathrm{E}}

\newcommand{\bb}{\mathbf{b}}

\newcommand{\vb}{\mathbf{v}}

\newcommand{\xb}{\mathbf{x}}
\newcommand{\yb}{\mathbf{y}}
\newcommand{\zb}{\mathbf{z}}

\newcommand{\Hb}{\mathbf{H}}
\newcommand{\Lb}{\mathbf{L}}
\newcommand{\Ib}{\mathbf{I}}
\newcommand{\Mb}{\mathbf{M}}
\newcommand{\Nb}{\mathbf{N}}
\newcommand{\Qb}{\mathbf{Q}}

\newcommand{\Fb}{\mathbf{F}_g}

\newcommand{\Pb}{\mathbf{P}}
\newcommand{\Sb}{\mathbf{S}}
\newcommand{\Kb}{\mathbf{K}}

\newcommand{\Cb}{\mathbf{C}}

\newcommand{\xl}{\tilde{\xb}}
\newcommand{\xhl}{\tilde{\xb}_c}
\newcommand{\vl}{\tilde{\vb}}

\newcommand{\xch}{\mathbf{\tilde{x}}_\perp}
\newcommand{\xpi}{\mathbf{\tilde{x}}_\parallel}
\newcommand{\sil}{\tilde{\boldsymbol{\sigma}}}

\newcommand{\sigmab}{\boldsymbol{\sigma}}

\title{\LARGE \bf
 Spatio-Temporal Communication Compression in Distributed Prime-Dual Flows
}

\author{Zihao Ren, Lei Wang, Deming Yuan, Hongye Su, Guodong Shi
\thanks{This research was supported by Zhejiang Provincial Natural Science Foundation of China under Grant No. LZ23F030008, and the National Natural Science Foundation of China under Grant No. 62203386.}
\thanks{Zihao Ren, Lei Wang and Hongye Su are with the College of Control Science and Engineering, Zhejiang University, P.R. China. (E-mail: zhren2000; lei.wangzju; hysu69@zju.edu.cn ).%
}
\thanks{Deming Yuan is  with the School of Automation, Nanjing University of Science and Technology,  P.R. China (Email: dmyuan1012@gmail.com).%
}
\thanks{Guodong Shi is with  Australia Centre for Field Robotics, The University of Sydney, Australia (Email: guodong.shi@sydney.edu.au).}
}

\begin{document}

\maketitle
\thispagestyle{empty}
\pagestyle{empty}


\begin{abstract}         

In this paper, we study distributed prime-dual flows for multi-agent optimization with spatio-temporal compressions. The central aim of multi-agent optimization is for a network of agents to collaboratively solve a system-level optimization problem with local objective functions and node-to-node communication by distributed algorithms. The scalability of such algorithms crucially depends on the complexity of the communication messages, and a number of communication compressors for distributed optimization have recently been proposed in the literature. First of all, we introduce a general spatio-temporal compressor characterized by the stability of the resulting dynamical system along the vector field of the compressor. We show that several important distributed optimization compressors such as the greedy sparsifier, the uniform quantizer, and the scalarizer all fall into the category of this spatio-temporal compressor. Next, we propose two distributed prime-dual flows with the spatio-temporal compressors being applied to local node states and local error states, respectively, and prove (exponential) convergence of the node trajectories to the global optimizer for (strongly) convex cost functions. Finally, a few numerical examples are present to illustrate our theoretical results.

\end{abstract}
{\bf keywords}:
Communication compression; distributed optimization; exponential convergence; compressors

\section{Introduction}\label{sec.Introduction}
Distributed intelligent systems, such as drone swarms, smart grids, and cyber-physical systems, have been extensively researched across disciplines such as control, signal processing, and machine learning \cite{magnusbook,martinez07,kar12,Rabbat2010}. The mathematical representation of a distributed system involves a network connecting multiple agents, where each node symbolizes an individual agent, and the edges depict communication lines between these nodes. 
When distributed systems are required to implement tasks such as cluster optimization and collaborative control, it requires the foundational functionality of distributed computing. In this process, each node stores localized information and communicates messages with connected nodes through the network, and collaboratively solves a mathematical problem \cite{magnusbook}. This paper focuses on addressing distributed optimization problems, where each node possesses a function, aiming to identify solutions that collectively minimize the sum of network node functions through constant communication across the network. The applications of distributed optimization span a wide range of domains \cite{NTAC,DQGO}.


Extensive research has been devoted to the study of distributed optimization algorithms, primarily rooted in the  consensus algorithm. The goal of this algorithm is to foster consistency in the states across nodes within the network \cite{SMAC,CCAO}. A combination of the consensus algorithm with the classical gradient descent method in optimization problems, coupled with stability tactics, results in the distributed subgradient algorithm (DSG), achieving sublinear convergence under a   strongly convex global cost function \cite{DSMF,DSCO}.
To address distributed optimization problems with faster rate requirements, more sophisticated algorithms have been introduced. The distributed gradient tracking algorithm (DGT) incorporates an additional state to trace the gradient of the objective function \cite{MB-GT,PGMF}, akin to integral action \cite{TGTI}. 
For diverse equivalent forms of distributed optimization problems, various Lagrangian functions have been proposed, giving rise to multiple algorithms based on the saddle point dynamic method. Examples include the Wang-Eila algorithm in \cite{JW-WE} and primal dual algorithm in \cite{XY-LCOF}, distinct in  communication states. 
Readers of interest are referred to \cite{SURV}  for more details of recent advances on distributed optimization algorithms.

In practical implementation, the network bandwidth for communication in distributed systems is limited and numerous communication compression strategies have been developed to handle such issues. 
In \cite{FCRO,CROD,FROD,LCOC}, the  idea of  quantifying the communication is combined with  DSG and DGT algorithms, {where
quantization can be regarded as a specific kind of compression}. Specifically, \cite{FCRO} applied random quantization and \cite{CROD} introduced adaptive quantization, where the quantizer codebook  changes when approaching the solution.  In \cite{LCOC}, the authors developed a dynamic encoding and decoding scheme for quantization.  In addition to quantization, there are also several other types of compressors capable of reducing communication bits by synthesizing concepts from quantization, sparsity, and randomization \cite{XY-CCFD,ALCA,DSOA,AR-AEQD}.
These compressors exclusively focus on the spatial dimension, encompassing the information within transmitted messages. Notably, the compressor in \cite{LW-DSFN} incorporates temporal dimensions, utilizing information across time. {\cite{ASCB,SCOT} also also adopt a combined spatial and temporal compression approach in areas outside of distributed computing. Although they do not compress communication messages between nodes in distributed networks, their approach is worth learning from.}

In addition,  how to combine the compressors with distributed optimization algorithms has become a noteworthy area of study. This is because refining the application method can facilitate the successful integration of more general compressors and enhance the overall effectiveness of the algorithm.
Beyond the direct application of compressors to the communication state, there exist intriguing techniques, as direct application often poses challenges to stability \cite{QIAF,DSMA}. For instance, \cite{AR-AEQD,CROD} incorporate a weighted sum of the updated value and the original value into the original value. {This approach ensures that the distortion caused by compression does not result in significant fluctuations with each iteration}. \cite{FCRO,DACW} compress the difference between iterations rather than the original value, {which ensures that each compressed value is small, avoiding large errors by the compressor when the state is large}. In the work of \cite{LCOC,DCWL}, the difference is scaled and then compressed, with the results communicated after a reverse reduction, further ensuring the convergence of the algorithm, {
ensuring that the fluctuations in the iteration values are not significant and allowing the compressor to fully capture the transmitted message.}.


{Our research also focuses on the above two aspects: a general compressor and its application methods. The former can help us find commonalities among different compressors. Moreover, we can determine whether a new specific compressor can be used and discover new specific compressors by it. The latter can help us find a general application method for this class of compressors without the need to analyze each specific compressor.}


In this paper,  we propose a type of ST compressors, which aims to compress the information from both the temporal and spatial perspectives and satisfies that the induced non-autonomous system is globally exponentially stable at the zero equilibrium.
Such ST compressors encompasses existing compressors in the literature, such as the scalarized compressors \cite{LW-DSFN} and the contractive compressors \cite{XY-CCFD}. Moreover, for a linear form of ST compressors, we prove that the compressors can be directly embedded into the prime-dual algorithm, and the resulting distributed compressed optimization algorithm can guarantee asymptotic (exponential) convergence to the global optimizer of (strongly) convex cost function. For nonlinear ST compressors, similar to \cite{XY-CCFD}, one can introduce a filtering step to derive a state difference for compression, leading to a distributed compressed prime-dual algorithm that can achieve the same convergence properties for (strongly) convex optimization problems.

The paper is structured as follows. Section \ref{sec.pro} formulates the distributed optimization problem of interest and proposes the spatio-temporal compressors for message communication. Our main work is in Section \ref{sec.mai}, where two communication-compressed algorithms are established on the prime-dual algorithm, and numerical simulations are presented to show the effectiveness of the proposed approaches in Section \ref{sec.num}. Finally, a brief conclusion is made in Section \ref{sec.con}.  All technical proofs are collected in the Appendices.

\emph{\bf Notation.}
In this paper, $\|\cdot \|$
denotes Euclidean norm. Assume $x$ is a vector of dimension $n$ and $A$ is a square matrix of dimension $n$, then $\|x\|^2_A$ denotes $x^TAx$.
The notation $\mathbf{1}_n(\mathbf{0}_n)$, $\mathbf{I}_n$ and $\{{\bf e}_1,...,{\bf e}_m\}$ denote the column one (zero) vector, identity matrix and base vectors in $\mathbb{R}^d$ respectively. The expression $diag(x_1,...,x_n)$ is a
diagonal matrix with the $i$-th diagonal element being $x_i$. The symbol $\otimes$ denotes the Kronecker product. For differential function, $\nabla(\cdot)$ denotes its gradient. 
\section{Problem Formulation}
\label{sec.pro}
\subsection{Distributed Optimization}

In this paper, we consider a network of agents indexed by $\mathrm{V}=\{1,2...n\}$, where each agent $i\in\mathrm{V}$ holds a cost function $f_i:\mathbb{R}^d\rightarrow \mathbb{R}$, and aims to solve the following distributed optimization problem 
\begin{equation}
    \label{eq:DO}
    \ba{c}
    \mathrm{min}\  \sum_{i=1}^n f_i(x_i)\\
    \mathrm{s.t.}\ x_i=x_j, \quad\forall i,j\in \mV.
    \ea
\end{equation}
Particularly, each local cost function $f_i$ is assumed to fulfill the following requirements.
\begin{assumption}\label{ass-DO}
    The following properties are satisfied.
    \begin{itemize}
    \item[i)] Each local cost function $f_i$, $i\in \mV$ is twice continuously differentiable, and its gradient  $\nabla f_i$ is globally Lipschitz continuous, satisfying $\|\nabla f_i(x_1)-\nabla f_i(x_2)\| \leq L_f\|x_1-x_2\|$ for any $x_1,x_2\in\mathbb{R}^d$ and some $L_f>0$. 
    \item[ii)] The global cost function $f(x):=\sum_{i=1}^n f_i(x)$ is (strongly) convex, i.e., for some $\mu=0$ ($\mu>0$), $f(x)$ satisfies
    $f(y)\geq f(x)+\nabla f(x)^T(y-x)+\frac{\mu}{2}\|y-x\|^2$ for all $x,y\in\mathbb{R}^d$. Moreover, $f(x)>-\infty$. \hfill$\square$ 
    \end{itemize}
\end{assumption}

\begin{remark}
   The twice continuous differentiability in Assumption \ref{ass-DO}.i)  is introduced to make the subsequent {proof progress} convenient, where the twice derivative  of $f_i$ is used. This assumption  in fact can be removed  when the forthcoming continuous-time algorithms are discretized. \hfill$\square$
\end{remark}

If Assumption \ref{ass-DO} holds, then the considered optimization problem \eqref{eq:DO} turns out a convex optimization problem, allowing an optimal solution set $\mathcal{S}\subseteq\mathbb{R}^d$ such that 
$\nabla f(s)=0$ and $f(s)=f^\ast$ for each $s\in \mathcal{S}$, where $f^\ast$ is the optimal value of \eqref{eq:DO}. If $\mu>0$, the set $\mathcal{S}$ has a unique element $s^\ast$.

As each agent has the information of only local cost function, to solve such a distributed optimization problem \eqref{eq:DO}, a communication network is usually required for transmitting messages. In this paper, we consider the communication graph $\mathrm{G=(V,E)}$, where $\mathrm E$ denotes the set of edges. 
Let $[a_{ij}]\in \mathrm{R}^{n\times n}$ denote the weight matrix complying with graph $\mathrm G$, i.e., $a_{ij}>0$ if $(j,i)\in\mE$ and $a_{ij}=0$ if $(j,i)\notin\mE$. Then denote by $L_G$  the Laplacian matrix of graph $\mathrm G$, satisfying $[L_G]_{ij}=-a_{ij}$ for all   $i\neq j$, and $[L_G]_{ii}=\sum_{j=1}^n a_{ij}$ for all $i\in\mathrm{V}$. 
For simplicity, it is assumed that the graph $\mG$ is undirected, connected and time-invariant, which implies that  the Laplacian matrix is symmetric and positive semi-definite, with eigenvalues $\lambda_i$, $i\in\mV$ in an ascending order satisfying $0=\lambda_1<\lambda_2\leq...\leq \lambda_n$ and $ L_{\mG}\mathbf{1}_n=\mathbf{0}_{n}$ by \cite{magnusbook}.

\subsection{Distributed Prime-Dual Algorithm}
With the above communication network $\mG$, several distributed optimization algorithms have been developed in the literature \cite{DSMF,DSCO,MB-GT,PGMF,TGTI,XY-LCOF,JW-WE} to compute the solution $s^\ast$ for  \eqref{eq:DO}. In this paper, we mainly focus on the distributed Primal-Dual flow, which enables to achieve exponential convergence and further generalizations to the case with optimization constraints \cite{DJ-PD1},\cite{PC-PD2}. A common distributed primal-dual flow for \eqref{eq:DO} takes the form \cite{XY-LCOF}
\begin{equation}
\ba{rcl}\label{eq:Prime_Dual}
\dot{x}_{i}&=&-\alpha\sum^{n}_{j=1}L_{ij} {x}_{j,c}-\beta v_{i}-\eta \nabla f_i(x_i) \\
\dot{v}_{i}&=&{\beta\sum^{n}_{j=1}L_{ij}{x}_{j,c}}\\
x_{i,c}&=&x_i,
\ea
\end{equation}
where $x_{i,c}$ denotes the communication messages over the network, and $\alpha, \beta,\eta>0$ are parameters to be fixed and the initial condition  $\sum_{i=1}^n v_i(0)=\mathbf{0}_d$.

It is clear from \eqref{eq:Prime_Dual} that the node state $x_i\in\mathbb{R}^{d}$ needs to be transmitted over the graph $\mG$ at each computation round. When the dimension $d$ is large, this usually means a heavy  communication bandwidth burden required for completing the computation task, and thus motivates to develop  compression (or quantization) strategies to reduce the communication bandwidth requirement, while preserving the exponential convergence.

\subsection{Communication Compressors}





In this paper, we are particularly interested in the spatio-temporal (ST) compressors which are characterized by the following definition.


\begin{definition}\label{def-C0}
    The compressor $\mathcal{C}(x_e,t):\mathbb{R}^d\times\mathbb{R}_+\rightarrow \mathbb{R}^d$ is a {\em ST compressor} if it satisfies the following two properties.
    \begin{itemize}
        \item[i).] The induced non-autonomous system 
    $\dot{x}_e=-\mathcal{C}(x_e,t)$ is uniformly globally exponentially stable at the zero equilibrium.
    \item[ii).] There exists a constant $L_c>0$ such that $\|\mathcal{C}(x_e,t)\| \leq L_c\|x_e\|$ for all $x_e\in\mathbb{R}^d$ and any $t\in\mathbb{R}_+$.\hfill$\square$
    \end{itemize}
\end{definition}

In the literature there are also some other classes of compressors. In the following we will show the connection between our ST compressors and the existing ones.

\begin{itemize}
    \item[1).] {\bf The scalarized compressor} $\mathcal{C}_1:\mathbb{R}^d\times\mathbb{R}_+\rightarrow \mathbb{R}^d$ satisfies $\mathcal{C}_1(x_e,t)=\psi(t)\psi(t)^Tx_e$, where the piece-wise continuous compression vector $\psi:\mathbb{R}_+\rightarrow \mathbb{R}^d$ is uniformly bounded and persistently excited, i.e., 
    \[
    \ba{l}
    \alpha_2 \Ib_d \geq  \int_{t}^{t+T_1} \psi(s)\psi^\top (s) ds \geq \alpha_1 \Ib_d\,,\quad \forall t\geq 0\,,
    \ea
    \]
    for some constants $\alpha_1,\alpha_2,T_1>0$ (see \cite{LW-DSFN}). 
    A specific example, denoted by $\mathcal{C}_{1a}$, can be derived by letting $\psi(t)= \mathbf{e}_i $ with $i=1+(k\   \mathrm{mod}\ d)$ for $t\in [k{\Delta}t,(k+1){\Delta}t), k\in\mathbb{N}$ and $\Delta t$ is step size. 


    \item[2).] {\bf The contractive compressor} $\mathcal{C}_2:\mathbb{R}^d\rightarrow \mathbb{R}^d$ satisfies 
\begin{equation}
    \label{ass_c2}
    \ba{rcl}
    \|\frac{\mathcal{C}_2(x_e)}{r}-x_e\|^2\leq (1-\varphi)\|x_e\|^2
    \ea
    \end{equation} 
    for some $\varphi\in(0,1]$ and $r>0$ (see \cite{XY-CCFD,YL-ACGT,AR-AEQD} with the expectation operator removed). By \cite{XY-CCFD}, the followings are specific examples of $\mathcal{C}_2$:
    \begin{itemize}
    \item[2a).] Greedy (Top-k) sparsifier \cite{Beznosikov2020Onbiased} 
$\mathcal{C}_{2a}(x_e)=\sum_{s=1}^{k}[x_e]_{i_s}{\bf e}_{i_s}$
where $i_1,...,i_k$ are the indices of 
largest $k$ coordinates in the absolute value of $x_e$.
\item[2b).] Standard uniform quantizer \cite{XY-CCFD}
$\mathcal{C}_{2b}(x_e)=\frac{\|x_e\|_\infty}{2}\mathrm{sgn}(x_e),$ 
where $\mathrm{sgn} (\cdot)$ denotes  the element-wise sign.
\end{itemize}

\end{itemize}

\begin{proposition}
The following statements are true.
\begin{itemize}
    \item[a).] Compressor $\mathcal{C}_1$ belongs to the ST compressor.
    \item[b).] Compressor $\mathcal{C}_2$ belongs to the ST compressor.\hfill$\square$
\end{itemize}

\end{proposition}
\begin{remark}
    In addition to the above mentioned compressors,  there are also some other forms of compressors that satisfy Definition \ref{def-C0}. For example, $\mathcal{C}_3(x_e,t)=\theta(t)\psi(x_e,t)$ where $\psi(x_e,t)$ is a scalarized mapping and $\theta(t)\psi(x_e,t)$ is strongly $P$-monotonic (see \cite{LW-RIIA}).\hfill$\square$
\end{remark}

\begin{remark}
    Taking practical implementation into account, we stress that when the compressor $\mathcal{C}$ is used, it does not mean to transmitting  the vector of $\mathcal{C}$.
    {
    We explain $\mathcal{C}_1$ and $\mathcal{C}_1$ separately. For $\mathcal{C}_1$, the sender transmits the scalar $\hat x_i(t)=\psi(t)^\top x_i(t)$, and then the receiver multiplies by  $\psi(t)^\top$ to obtain $\mathcal{C}_1(x_i,t)$. For $\mathcal{C}_2$, taking $\mathcal{C}_{2b}$ as an example, the sender transmits the information $\|x_i\|_\infty$ and a vector $\mathrm{sgn}(x_i)$, which requires fewer bits compared to the original vector $x_i$, and the receiver uses the prior information of $\mathcal{C}_{2b}$ to complete the recovery.}
    \hfill$\square$
\end{remark}

\begin{remark}
In contrast with the conventional compressors, e.g., the contractive compressor { (though it belongs to the ST compressor)}, the ST compressor exhibits two distinctive features. Firstly, it synthesizes information from both the time and space domains, broadening its applicability and expanding the design possibilities. Secondly, its key characteristic is elucidated through a non-autonomous system, which can simplify the design procedure while  providing the flexibility to incorporate control-related tools into distributed optimization.
\end{remark}

\section{Main results}
\label{sec.mai}
In this section, we propose distributed compressed optimization flows over the  graph $\mathrm{G}$.

\subsection{Distributed Prime-Dual Flow with Direct State Spatio-Temperal Compressor (DPDF-DSSTC)}


An intuitive design of compressed optimization algorithm is to replace the transmitted message $x_i$ by the compressed  one, i.e., $x_{i,c}=\mathcal{C}(x_i,t)$ in \eqref{eq:Prime_Dual}, leading to the following DPDF-DSSTC

\begin{equation}
\ba{rcl}\label{eq:Algorithm1}
\dot{x}_{i}&=&-\alpha\sum^{n}_{j=1}L_{ij} {x}_{j,c}-\beta v_{i}-\eta \nabla f_i(x_i) \\
\dot{v}_{i}&=&{\beta\sum^{n}_{j=1}L_{ij}{x}_{j,c}}\\
x_{i,c}&=&\mathcal{C}(x_i,t),
\ea
\end{equation}
with  initial conditions $x_i(0)\in\mathbb{R}^d$ and $\sum_{i=1}^n v_i(0)=\mathbf{0}_d$.

{
Due to the information distortion caused by the compressors, the system often requires additional states to eliminate the positive definite error term between the communication message $x_{i,c}(t)$ and the original state $x_{i}(t)$.
As shown in \cite{QIAF,DSMA},  almost every kind of compressors lead to the challenge to stability when applied directly, including ST compressors. Luckily, we find that these risks can still be avoided if certain additional conditions are satisfied. 
We are ready to propose the following theorem for  \eqref{eq:Algorithm1}.}

\begin{theorem}\label{thm-1}
Let Assumption \ref{ass-DO} hold, and the ST compressor $\mathcal{C}(x_e,t)$ be linear in $x_e$, e.g., in the form of the scalarized compressor. Then for the  DPDF-DSSTC  \eqref{eq:Algorithm1}  with constants $\alpha,\beta,\eta>0$ in Table \ref{tab:the1}, the following statements are true.
\begin{itemize}
    \item[i)] There exists $s \in\mathcal{S}$ such that $\lim_{t\to\infty}x_i(t)= s$.
    \item[ii)] Suppose $f(x)$ is strongly convex, i.e., $\mu>0$ in Assumption \ref{ass-DO}, then DPDF-DSSTC  \eqref{eq:Algorithm1} converges to the optimal solution $s^\ast$ exponentially, i.e.,
\[
\ba{rcl}
\|x_i(t) - s^\ast\|^2 = \mathcal{O}(e^{-\gamma t})
\ea
\]
for some $\gamma>0$ (see Table \ref{tab:the1} for upper bound of the parameters).\hfill$\square$
\end{itemize}

\end{theorem}

\begin{table}[htbp]
    \centering
    \begin{tabular}{c|c||c|c}
    \hline
        $\alpha$ & =1 &$\eta$&$\mathrm{min}\{\beta^5,\beta,\frac{\xi_3}{4\xi_4},1,\sqrt{\frac{3}{8\xi_{6}}}\}
        $\\
        \hline
        $\beta$ & $ \mathrm{min}\{\frac{c_3}{3\xi_1'}, \frac{c_3r}{3\xi_2},\frac{c_1}{2}\}$&$\gamma$ & $\mathrm{min}\{\frac{c_3}{3\beta},\xi_3\beta^2,\eta\frac{\mu}{2n}\}$\\
       
        \hline
    \end{tabular}
    \caption{The range of parameters of Theorem \ref{thm-1} (see Appendix \ref{pr:The1} for values of $\xi_1',\xi_2,\xi_3,\xi_4,\xi_6,c_1,c_3,r$).}
 \label{tab:the1}
\end{table}



\subsection{Distributed Prime-Dual Flow with Error State Spatio-Temperal Compressor (DPDF-DSETC)}

In the previous subsection, we have shown that the linear ST compressor can be directly applied into the flow \eqref{eq:Prime_Dual} by modifying $x_{i,c}=\mathcal{C}(x_i,t)$, while maintaining the asymptotic (exponential) convergence when solving a (strongly) convex optimization problem.
However, we note that for other more general types of the ST compressors, such a direct application may fail to maintain the convergence property.
Particularly, the instability may arise due to the fact that the compressed value $x_{i,c}(t)$ in Flow \eqref{eq:Algorithm1} is both nonzero and time-varying, posing challenges when employing nonlinear compressors.

In the subsequent context, inspired by \cite{XY-CCFD} we show that by introducing a distributed filter and a distributed integrator and compressing the state errors, the aforementioned challenge can be overcome, leading to a general distributed optimization framework for the ST compressors. The proposed DPDF-DSETC is given below.
\begin{equation}
\ba{rcl}\label{eq:Algorithm2_im}
\dot{\sigma}_i&=&q_i \\
\dot{z}_i&=&q_i-\sum^{n}_{j=1}L_{ij}q_{j}\\
\dot{x}_{i}&=&-\alpha(\sigma_i-z_i+\sum^{n}_{j=1}L_{ij}{q}_{j})-\beta v_{i}-\eta \nabla f_i(x_i) \\
\dot{v}_{i}&=&{\beta(\sigma_i-z_i+\sum^{n}_{j=1}L_{ij}{q}_{j})}\\
q_i &=& \mathcal{C}(x_i-\sigma_i,t),
\ea
\end{equation}
where $\sigma_i\in\mathbb{R}^d$ and $z_i\in\mathbb{R}^d$ are extra states of the introduced filter and integrator, respectively. The initial condition is $\sum_{i=1}^n v_i(0)=\mathbf{0}_d$ and $\sigma_i(0)= z_i(0)=\mathbf{0}_d,\forall 
i\in \mV$. It is worth noting that for Flow \eqref{eq:Algorithm2_im}, the communication message is the compressed version of the error between states $x_i$ and $\sigma_i$.
We are now ready to present the following results for Flow \eqref{eq:Algorithm2_im}.

\begin{theorem}\label{thm-2}
Let Assumption \ref{ass-DO} hold, and $\mathcal{C}(x_e,t)$ is a ST compressor. Then for constants $\alpha,\beta,\eta>0$ in Table \ref{tab:the2}, the DPDF-DSETC  \eqref{eq:Algorithm2_im} with  a ST compressor $\mathcal{C}(x_e,t)$ satisfy the following statements.
\begin{itemize}
    \item[i)] There exists $s \in\mathcal{S}$ such that $\lim_{t\to\infty}x_i(t)= s$.
    \item[ii)] Suppose $f(x)$ is strongly convex, i.e., $\mu>0$ in Assumption \ref{ass-DO}, then the DPDF-DSETC  \eqref{eq:Algorithm2_im}  converges to the optimal solution $s^\ast$ exponentially, i.e.,
\[
\ba{rcl}
\|x_i(t) - s^\ast\|^2 = \mathcal{O}(e^{-\gamma t})
\ea
\]
for some $\gamma>0$ (see Table \ref{tab:the2} for upper bound of the parameters).\hfill$\square$
\end{itemize}

\end{theorem}

\begin{table}[htbp]
    \centering
    \begin{tabular}{c|c||c|c}
    \hline
        $\alpha$ & $ \mathrm{min}\{\frac{c_3r}{3\xi_8}, \frac{c_3}{3\xi_9}\}$ &$\eta$&$\mathrm{min}\{\beta^5,\beta,\frac{1}{\alpha^{2}},1,\frac{\xi_1}{8\xi_2'},\frac{\xi_5}{4\xi_6},{\frac{3}{8\xi_{10}}}\}$\\
        \hline
        $\beta$ & $\mathrm{min}\{\frac{\alpha}{2},\frac{\xi_1\alpha}{4\xi_4}\}$ &$\gamma$&$\mathrm{min}\{\frac{\xi_1 \alpha}{2},\frac{\xi_5\beta}{2},\frac{c_3}{3c_1},\eta\frac{\mu}{2n}\}$\\
        \hline
    \end{tabular}
    \caption{The range of parameters of Theorem \ref{thm-2} (see Appendix \ref{pr:The2} for values of $\xi_1,\xi_2',\xi_4,\xi_5,\xi_6,\xi_8,\xi_9,\xi_{10},c_3,r$).}
 \label{tab:the2}
\end{table}

\begin{remark}
We stress that proposed the ST compressors can  be  be combined with other types of distributed optimization algorithms, such as distributed subgradient algorithm, gradient tracking algorithm in \cite{MB-GT} and Wang-Elia
algorithm in \cite{JW-WE}  using the same techniques as in flow \eqref{eq:Algorithm2_im}, and the same conclusion can be concluded as Theorems \ref{thm-1} and \ref{thm-2}. \hfill$\square$
\end{remark}

\section{Numerical Simulations}
\label{sec.num}

In this section,
numerical simulations are presented to verify the validity of the proposed algorithms with compressors. 


We consider a network of $n$ nodes over a
circle communication graph and dimension of local state is $d$, where each edge is assigned with the same unit weight and each node holds a local function $f_i(x_i)=\frac{1}{2}\|\Hb_i x_i-b_i\|^2 $ with some randomly generated $\mathbf{H}_i\in\mathbb{R}^d$ and $b_i\in\mathbb{R}$. Assume that the linear equation $\Hb x=\bb$ has a unique solution 
$s^\ast$, where $\mathbf{H}= [\mathbf{H}_1 \,\dots  \, \mathbf{H}_n]^\top\in\mathbb{R}^{n\times d}$ and $\mathbf{b}=[b_1\,\dots \,b_n]^\top\in\mathbb{R}^{d}$,  then we can conclude that the functions $f_i(x_i)$ satisfy Assumption \ref{ass-DO} with $\mu>0$ and optimal solution $s^\ast$.
Specifically, we let $n=10$, $d=5$ and $s^\ast= [1,3,-1,4,2]$.

We initially implement Flow \eqref{eq:Algorithm1} to address the distributed optimization problem. In this application, we integrate compressor $\mathcal{C}_{1a}$ into Flow \eqref{eq:Algorithm1} using a step size of $\Delta t=0.01$ and the parameters $\alpha=1$, $\beta=0.5$, $\eta=0.1$. The plot illustrates the sum of squared distances from the current $x_i(t)$ to $s^\ast$, denoted as $\sum_{i=1}^n\|x_i(t)-s^\ast\|^2$ over time in Flow \eqref{eq:Algorithm1}. Notably, Flow \eqref{eq:Algorithm1} exhibits exponential convergence to the optimal solution, verifying Theorem \ref{thm-1}. { At the same time, we can see that $\mathcal{C}_{2a}$ and $\mathcal{C}_{2b}$, which do not satisfy the linear conditions, cause the system to lose convergence}.



We proceed to implement Flow \eqref{eq:Algorithm2_im} and analyze the results. In this application, we integrate compressors $\mathcal{C}_{1a}$, $\mathcal{C}_{2a}$ and $\mathcal{C}_{2b}$ into Flow \eqref{eq:Algorithm2_im} with $k=2$, a step size of $\Delta t=0.01$, and specific parameters $\alpha=1$, $\beta=0.5$, $\eta=0.1$. The evolution of the value $\sum_{i=1}^n\|x_i(t)-s^\ast\|^2$ over time in Flow \eqref{eq:Algorithm2_im} is shown. Notably, Flow \eqref{eq:Algorithm2_im} demonstrates exponential convergence to the optimal solution, aligning with the conclusion of Theorem \ref{thm-2}.


\section{Conclusions}
\label{sec.con}
In this paper, we have introduced a type of spatio-temporal compressors that integrates both spatial and temporal characteristics, effectively compresses information by leveraging information from both the time and space domains. This class of compressors has covered several assumptions in literature on compressors. Our proposed compressor has been implemented in two distinct compression algorithms based on the primal-dual algorithm. In the future, we will investigate a broader spectrum of compressor types or enhanced algorithms tailored to the characteristics of this compressor, and to have extended its application to more classical distributed optimization algorithms, examining its universality across different algorithms.

\appendix

\subsection{Proof of Proposition 1} 
\label{pr:dis}

\emph{Proof of a).}  We proceed by showing the compressor $\mathcal{C}_1(x_e,t)=\psi(t)^T\psi(t)x_e$ satisfies Properties i) and ii) of Definition \ref{def-C0}, respectively. The proof of the first property is obvious by recalling \cite{Brian-TAC-1977} that system $\dot{x}_e=-\psi(t)^T\psi(t)x_e$ is globally exponentially stable at the zero equilibrium if and only
if $\psi$ is PE. The second can be shown by noting that $\psi(t)$ is uniformly bounded.

\emph{Proof of b).} Similarly, we proceed to show the compressor $\mathcal{C}_2(x_e)$ satisfying both Properties of Definition \ref{def-C0}.
Note that the contractive compressor \eqref{ass_c2} is equivalent to
\begin{equation}
\ba{rcl}
\label{ass_c2'} \|{\mathcal{C}_2(x_e)}/{r}\|^2-2{x_e^T\mathcal{C}_2(x_e)}/{r}\leq -\varphi\|x_e\|^2\,.
\ea
\end{equation}

First, we prove that the system 
    $\dot{x}_e=-\mathcal{C}_2(x_e,t)$ is exponentially stable at the zero equilibrium. By choosing the Lyapunov function $V_e(x_e)={\|x_e\|^2}/{r}$ and using \eqref{ass_c2'}, we have
\[
    \ba{rcl}
\dot{V}_e=-2\frac{x_e^T\mathcal{C}_2(x_e)}{r}\leq -\varphi\|x_e\|^2-\|{\mathcal{C}_2(x_e)}/{r}\|^2.
    \ea
\]
Thus $x_e$-system is globally exponentially stable at the zero equilibrium and the Property i) is proved.

Then, by \eqref{ass_c2'} and using the Young's inequality, we have
\begin{equation}
\label{eq:C2_1}
    \ba{rcl}
\|{\mathcal{C}_2(x_e)}/{r}\|^2&\leq&  \frac{1}{2}\|{\mathcal{C}_2(x_e)}/{r}\|^2-(\varphi-2)\|x_e\|^2\\
\Rightarrow
\|{\mathcal{C}_2(x_e)}\|&\leq&  r\sqrt{{2(2-\varphi)}}\|x_e\|\leq 2r\|x_e\|,
    \ea
\end{equation}
where the last inequality is obtained by $\varphi\in(0,1]$. Thus the Property ii) is proved with $L_c=2r>0$. This completes the proof.




\subsection{Proof for Theorem 
\ref{thm-1}}
\label{pr:The1}


As $\mathcal{C}(x_e,t)$ is linear for $x_e$, for convenience, we let $\mathcal{C}(x_e,t)=\mathcal{A}(t)x_e$, where $\|\mathcal{A}(t)\|$ has a uniformly upper bound $a_m$ obtained from {the Property ii) of the ST compressor}. We let $\alpha=1$,  then Flow   \eqref{eq:Algorithm1} can be written in a tight form as 
\begin{equation}
\ba{rcl}\label{eq:Algorithm1_tight}
\dot{\xb}&=&-[ \Lb{\xb_c}+\beta \vb+\eta \Fb(\xb)] \\
\dot{\vb}&=&{\beta \Lb\xb_c}\\
\xb_c&=&(\mathbf{I}_n\otimes \mathcal{A}(t))\xb,\\

\ea
\end{equation}
where $\xb(t):=[x_1^T(t),...x_n^T(t)]^T
$, $\vb(t):=[v_1^T(t),...v_n^T(t)]^T
$, $\Fb(\xb):=[\nabla f_1^T(x_1)...\nabla f_n^T(x_n)]^T
$ and $\Lb:=L_G\otimes \mathbf{I}_d$.

As $f(x)$ is convex, there exists some $s^\ast\in\mathbb{R}^d$ that $\nabla f(s^\ast)=0$. Then define the state error $\tilde{\xb}(t):=\xb(t)-(\mathbf{1}_n\otimes \mathbf{I}_d)s^\ast$, $\tilde{\vb}(t):=\vb(t)+\frac{\eta \Fb(\Hb\xb(t))}{\beta}$, where $\Hb:=\frac{1}{n}\mathbf{1}_n\mathbf{1}_n^T\otimes \mathbf{I}_d$. 
Taking the time derivative of the state errors along \eqref{eq:Algorithm1_tight} yields
\begin{equation}
\ba{rcl}\label{eq:Algorithm1_tight_0}
\dot{\xl}&=&-[ \Lb\xhl+\beta \vl+\eta [\Fb(\xb)-\Fb(\Hb\xb)]] \\
\dot{\vl}&=&{\beta \Lb\xhl}+\frac{\eta}{\beta}\dot{\Fb}(\Hb\xb)\\
\xhl&=&(\mathbf{I}_n\otimes \mathcal{A}(t))\xl.
\ea
\end{equation}
As $\Hb\vb(0)=\mathbf{0}_{nd}$, and noting that \begin{equation}
\ba{c}\label{eq:HL}
\Hb\Lb=\Lb\Hb=\mathbf{0},
\ea
\end{equation}
we can conclude that $\Hb\vb=\mathbf{0}_{nd}$, and thus
\begin{equation}
\label{eq:Hv}
\ba{rcl}
\Hb(\vl(t)-\eta \Fb(\Hb\xb(t))/\beta)=\mathbf{0}_{nd}\,,\quad \forall t\in\mathbb{R}_+.
\ea
\end{equation}

We further define $\xch:=\Kb\xl$ and $\xpi:=\Hb\xl$, 
where $\Kb=\Sb\Sb^T$, $\Sb=S\otimes \mathbf{I}_d$ with  $S\in\mathbb{R}^{n\times(n-1)}$ being a matrix whose rows being eigenvalue vectors corresponding to nonzero eigenvalues of 
$L_G$,  satisfying
\begin{equation}\label{eq:HK}
\ba{rcl}
S^\top\mathbf{1}_n =\mathbf{0}_{n-1}\quad  \mathbf{I}_n = SS^\top +\mathbf{1}_n\mathbf{1}_n^\top/n .
\ea
\end{equation}
It is clear that $\xl= \xch + \xpi$.

By \eqref{eq:HL}, \eqref{eq:HK} and \begin{equation}
\ba{c}\label{eq:KL}
\Kb\Lb=\Lb\Kb=\Lb,
\ea 
\end{equation}
the system \eqref{eq:Algorithm1_tight_0} can be further transformed by 
\begin{equation}
\ba{rcl}\label{eq:Algorithm1_tight_0_f}
\begin{bmatrix}
    \mathbf{\dot{\tilde{x}}}_\perp\\
    \dot{\vl}
\end{bmatrix}&=&\Nb
\begin{bmatrix}
    \mathbf{\xhl}\\
    {\vl}
\end{bmatrix}+\Mb(\xb)
\\
\mathbf{\dot{\tilde{x}}}_\parallel&=&- \Hb\eta \Fb(\xb) \\
\xhl&=&(\mathbf{I}_n\otimes \mathcal{A}(t))\xl,
\ea
\end{equation}
where $\Nb=\begin{bmatrix}
    -\alpha \Lb &-\beta \mathbf{I}_{nd}\\
    \beta \Lb &\mathbf{0}_{nd}
\end{bmatrix}$, $\Mb(\xb)=\begin{bmatrix}
    \eta [\Kb\Fb(\xb)-\Fb(\Hb\xb)]]\\
    \frac{\eta}{\beta}\dot{\Fb}(\Hb\xb)&
\end{bmatrix}$.

Before we study the stability of system \eqref{eq:Algorithm1_tight_0_f}, in the following some properties of the mapping $\Fb(\cdot )$ are presented.
With Assumption \ref{ass-DO}
and system \eqref{eq:Algorithm1_tight_0}, we conclude that 

\setstretch{0.5}
\begin{equation}\label{eq:G_yz}
\ba{rcl}
   \|\Fb(\yb)-\Fb(\zb)\| \leq L_f\|\yb-\zb\| \quad \forall \yb,\zb\in\mathbb{R}^{nd}\,
   \ea
  \end{equation}
\begin{equation}
    \ba{rcl}
    \label{eq:G_xHx}
&&\|\Fb(\xb)-\Fb(\Hb\xb)\|_\Kb^2\leq\|\Fb(\xb)-\Fb(\Hb\xb)\|^2\\
&\leq& L_f^2\|\xb\|_K^2=L_f^2\|\xch\|^2
\ea
\end{equation}
\begin{equation}
 \label{eq:G_dHx}
\ba{rcl}
&&\|\dot{\Fb}(\Hb\xb)\|^2=\|\frac{\partial \Fb}{\partial \Hb\xb}\Hb\dot{\xl}\|^2\leq \eta^2 L_f^2 \|\Hb\Fb(\xb)\|^2\\
&\leq &\eta^2L_f^4(\|\xch\|^2+\|\xpi\|^2),
\ea
\end{equation}
\setstretch{1}
where the second equality is obtained by \eqref{eq:HK} 
and the last equality is obtained by 
\begin{equation}
\ba{rcl}
    \label{eq:HG}\Hb\Fb((\mathbf{1}_n\otimes \mathbf{I}_d)s^\ast)=\mathbf{0}_{nd},
    \ea
\end{equation}
which is derived from the convexity of $f(x)$ in Assumption 
\ref{ass-DO}, \eqref{eq:HL} and \eqref{eq:HK}.

With such properties in mind, we choose $V_1(\xch,\vl)=\|\begin{array}{c}
    \xch\\
    \vl
\end{array}\|^2_\Qb$,
where we define $\Qb:=\frac{1}{2}\begin{bmatrix}
    \mathbf{0}_{nd} & \Pb\\
    \Pb & \frac{\Pb}{\beta}
\end{bmatrix}$, with $\Pb := \begin{bmatrix}
    \mathbf{1}_n/\sqrt{n}  &  S \\
\end{bmatrix}
   \begin{bmatrix}
    \lambda_n^{-1} &  \\
     & \Lambda^{-1}
\end{bmatrix}
\begin{bmatrix}
   \mathbf{1}_n^T/\sqrt{n}  \\ S^T
\end{bmatrix}\otimes \mathbf{I}_d$, for $\Lambda = \mathrm{diag}
 (\lambda_2,...,\lambda_n)
$. It can be easily concluded that 
\begin{equation}
\ba{c}\label{eq:PI}\lambda_n^{-1}\mathbf{I}_{nd}\leq \Pb\leq \lambda_2^{-1}\mathbf{I}_{nd}
\ea
\end{equation}
\begin{equation}
\ba{c}\label{eq:PL}
\Pb\Lb=\Lb\Pb=\Kb.
\ea
\end{equation}
\begin{equation}
\ba{c}
\Nb^T\Qb+\Qb\Nb=-\beta\begin{bmatrix}
    -\Kb & \mathbf{0}_{nd} \\
    \mathbf{0}_{nd} & \Pb

\end{bmatrix}.
\ea
\end{equation}
Then, computing the time-derivative of $V_1$ along \eqref{eq:Algorithm1_tight_0_f} yields
\begin{equation}
\label{eq:V1a_1}
\ba{rcl}
\dot{V}_1  
&=&  \beta \xch^T  \xhl+\frac{1}{\beta}\frac{\eta}{\beta}\dot{\Fb}(\Hb\xb)^T\Pb\vl+\frac{\eta}{\beta}\xl^T 
    \Kb\Pb\dot{\Fb}(\Hb\xb)\\
    &&-\vl^T \Pb [\beta \vl-\eta \Fb(\Hb\xb)+\eta \Fb(\xb) -\eta \Hb\Fb(\xb)]\\
&\leq&
(\beta a_m+\frac{\eta}{2}+\frac{\eta L_f^2}{8})\|\xch\|^2-(\beta-\frac{4\eta}{\lambda_2}-\frac{\eta\beta}{4})\|\vl\|_\Pb^2\\&&+(\frac{\eta^3 L_f^2}{2\beta^2\lambda_2^2}+\frac{1}{8}\eta+\frac{\eta^3L_f^2}{\lambda_2\beta^5})\|\Hb\Fb(\xb)\|^2,
\ea
\end{equation}
where the first equality is obtained by \eqref{eq:HK}, \eqref{eq:PL} and the fact $\Kb\xch=\xch$, and the first inequality is obtained by \eqref{eq:G_xHx}, \eqref{eq:PI} and the fact
\[
\ba{rcl}
\xch^T\xhl=\xl^T(SS^T\otimes\mathcal{A}(t))\xl\leq a_m\|\xch\|^2.
\ea
\]

By definition of $\mathcal{C}(x_e,t)$, it is easy to find that the system 
\[
\ba{rcl}
\dot{\zb}_e=-\Lambda \otimes \mathcal{A}(t) \zb_e,\quad \zb_e\in\mathbb{R}^{(n-1)d}
\ea
\] 
is exponentially stable at the zero equilibrium. By recalling the converse Lyapunov Theorem for exponential stability\cite[Theorem 4.14]{Khalil(2002)}, this implies the existence of a Lyapunov function $V_e:\mathbb{R}^{(n-1)d}\times\mathbb{R}_+\rightarrow\mathbb{R}_+$ which satisfies 
\begin{equation}
\ba{l}\label{eq:Ve_1}
c_1\|\zb_e\|^2 \leq V_{e}(\zb_e,t)\leq c_2\|\zb_e\|^2\\
  \frac{\partial V_{e}}{\partial t} - \frac{\partial V_{e}}{\partial \zb_e} (\Lambda\otimes\mathcal{A}(t))(\zb_e) \leq -c_3\|\zb_e\|^2\\\
  \|\frac{\partial V_{e}}{\partial \zb_e}\| \leq c_4 \|\zb_e\|
\ea
\end{equation}
for some $c_1,c_2,c_3,c_4>0$.
Thus, by choosing $V_2(\Sb^T\xl,t):=V_e(\Sb^T\xl,t)$, we have
\begin{equation}
\ba{rcl}\label{eq:V2_1}
\dot{V}_2 &=& \frac{\partial V_{2}}{\partial t} + \frac{\partial V_{2}}{\partial (\Sb^T\xl)} (\Sb^T\dot{\xl})\\ 
&\leq& -c_3 \|\xch\|^2 + c_4\|\xl\|\| \beta \vl+\eta[\Fb(\xb)-\Fb(H\xb)]\| \\
&\leq& -(c_3-
c_4\beta/r-c_4\eta/r) \|\xch\|^2 \\
&&+ c_4\beta r\lambda_n\|\vl\|_\Pb^2+c_4\eta rL_f^2\|\xch\|^2,
\ea
\end{equation}
where the first inequality is obtained by \eqref{eq:Ve_1} and the last inequality is obtained by \eqref{eq:PI}, \eqref{eq:G_xHx} and Young's Inequality, where $r>0$ is a  parameter to be determined later.

Before we proceed to the proof of statements i) and ii),  in the following we introduce some useful parameters, independent of $\beta$, $r$ and $\eta$.
\[
\ba{rcl}
&&\xi_0=\frac{2L_f^4}{\lambda_2\mu_n}+\frac{L_f^4}{\lambda^2_2\mu_n}+\frac{L_f^2}{4\mu_n},\\
&&\xi_{1}=a_m+\frac{1}{2}+\frac{5L_f^2}{8},\quad\xi_{2}=2c_4,\quad \xi_{3}=\frac{3}{4},\\
&&\xi_{4}=\frac{4}{\lambda_2}, \quad \xi_{5}=c_4\lambda_n,\quad \xi_{6}=\frac{L_f^2}{2\lambda_2^2}+\frac{L_f^2}{\lambda_2},\\
&&\xi_{1}'=\xi_{1}+c_4L^2_f+\frac{\xi_0L_f^2}{\mu_n}+\frac{L_f^2}{8}+\frac{L^4_f}{2\lambda^2_2}+\frac{L_f^4}{\lambda_2}.
\ea
\]

\emph{Proof of i).}
Choose $V_{3,a}(\xl)=f(\frac{1}{n}((\mathbf{1}_n^T\otimes\mathbf{I}_d)\xl))-f^\ast$, then
\begin{equation}
    \label{eq:V3a_1}
    \ba{rcl}
    \dot{V}_{3,a}
    &=&-\eta \|\Hb\Fb(\xb)\|^2+\eta\Fb^T(\xb)\Hb(\Hb\Fb(\xb)-\Fb(\Hb\xb))\\
    &\leq& -\frac{\eta}{2} \|\Hb\Fb(\xb)\|^2+\frac{\eta}{2}L^2_f\|\xch\|^2,
    \ea
\end{equation}
where the first equality is obtained by \eqref{eq:Algorithm1_tight} and the first inequality is obtained by \eqref{eq:G_xHx}.

To this end, we choose  $V(\xl,\vl)= V_1(\Kb\xl,\vl)+V_2(\Sb^T\xl,t)+V_{3,a}(\xl)$. With \eqref{eq:Ve_1}, it is easy to prove that
$V$ satisfies 
\begin{equation}
\ba{rcl}\label{eq:V_1a}
V\geq \beta\|\xch\|^2+\frac{1}{4\beta}\|\vl\|^2_\Pb
\ea
\end{equation}
as we choose $\beta\leq\frac{c_1\lambda_2}{2}$.

By fixing
\[
\ba{rcl}
 &&r=\mathrm{min}[\frac{\xi_3}{2\xi_5},1],\ \alpha=1,\ \beta\leq \mathrm{min}[\frac{c_3}{3\xi_1}, \frac{c_3r}{3\xi_2},\frac{c_1\lambda_2}{2}],\\ &&\eta\leq \mathrm{min}[\beta^5,\beta,\frac{\xi_3\beta}{4\xi_4},1,\sqrt{\frac{3}{8\xi_{6,a}}}],
 \ea
\]
and using \eqref{eq:V1a_1}, \eqref{eq:V2_1} and \eqref{eq:V3a_1}, we have
\[
    \ba{rcl}
    \dot V&\leq& -\frac{c_3}{3}\|\xch\|^2-\frac{\xi_3\beta}{4}\|\vl\|_\Pb^2.
    \ea
\]


With \eqref{eq:V_1a} and 
Invariance-like Theorem \cite[Theorem 8.4]{Khalil(2002)}, we have $\lim_{t\to \infty}\|\xch(t)\|=\lim_{t\to \infty}\|\vl(t)\|=0$. Then $\lim_{t\to \infty}\Hb\Fb(\xb(t))=0$ by \eqref{eq:Algorithm1_tight}, which implies $\lim_{t\to \infty}\xb(t)=(\mathbf{1}_n\otimes \mathbf{I}_d)s^\ast$ with the convexity of $f(x)$. The prove of Theorem 
\ref{thm-1}.i) is complete.

\emph{Proof of ii).} As $f(x)$ is strongly convex with $\mu>0$, we can derive 
\begin{equation}
\ba{rcl}
    \label{eq:G_mu}
    &&\xl^T\Hb(\Fb(\Hb\xb)-\Fb((\mathbf{1}_n\otimes \mathbf{I}_d)s^\ast))\\
    &=&\frac{1}{n}((\mathbf{1}_n^T\otimes\mathbf{I}_d)\xl)^T[\nabla f(\frac{1}{n}((\mathbf{1}_n^T\otimes\mathbf{I}_d)\xb))-\nabla f(s^\ast)]\\&\geq& \mu_n \|\xpi\|^2,
    \ea
\end{equation}
where $\mu_n:=\frac{\mu}{n}$. 

By \eqref{eq:G_dHx} and \eqref{eq:V1a_1},
\begin{equation}
\label{eq:V1b_1}
\ba{rcl}
\dot{V}_1  
&\leq&
(\frac{\eta^3L_f^4}{\lambda_2\beta^5}+\beta a_m+\frac{\eta}{2}+\frac{\eta L_f^2}{8}+\frac{\eta^3 L_f^4}{2\beta^2\lambda_2^2}+\frac{L_f^2}{8}\eta)\|\xch\|^2\\
    &&-(\beta-\frac{4\eta}{\lambda_2}-\frac{\eta\beta}{4})\|\vl\|_\Pb^2\\&&+(\frac{\eta^3 L_f^4}{2\beta^2\lambda_2^2}+\frac{L_f^2}{8}\eta+\frac{\eta^3L_f^4}{\lambda_2\beta^5})\|\xpi\|^2,
\ea
\end{equation}

Define $V_{3,b}(\xpi):=\frac{1}{2}\|\xpi\|^2$, then 
\begin{equation}
    \label{eq:V3b_1}
    \ba{rcl}
    \dot V_{3,b}
    &=& -\eta \xl^T\Hb(\Fb(\xb)-\Fb(\Hb\xb)\\
    &&+\Fb(\Hb\xb)-\Fb((\mathbf{1}_n\otimes \mathbf{I}_d)s^\ast))\\
    &\leq& -\eta \mu_n \|\xpi\|^2+\eta \|\xpi\| L_f \|\xch\|\\
    &\leq& -\eta\frac{\mu_n}{2} \|\xpi\|^2+\eta\frac{1}{2\mu_n}L_f^2 \|\xch\|^2,
    \ea
\end{equation}
where the first equality is obtained by \eqref{eq:Hv}, the first inequality is obtained by \eqref{eq:HG} and the second inequality is obtained by \eqref{eq:HK}, \eqref{eq:G_yz} and \eqref{eq:G_mu}.

We define the Lyapunov functions of system \eqref{eq:Algorithm1_tight_0_f} $ V(\xl,\vl): = V_1(\Kb\xl,\vl)+V_2(\Sb^T\xl)+2\xi_0V_{3,b}(\Hb\xl)$. As we choose $\beta\leq\frac{c_1\lambda_2}{2}$, it is easy to prove that $V$ is positive definite. In fact, 
\begin{equation}
\ba{rcl}\label{eq:V_1b}
V\geq\beta\|\xch\|^2+\frac{1}{4\beta}\|\vl\|^2_\Pb+ \xi_0\|\xpi\|^2.
\ea
\end{equation}



By fixing
\[
\ba{rcl}
&&r=\mathrm{min}[\frac{\xi_3}{2\xi_5},1], \alpha=1, \beta\leq \mathrm{min}[\frac{c_3}{3\xi_1'}, \frac{c_3r}{3\xi_2},\frac{c_1}{2}],\\
&&\eta\leq \mathrm{min}[\beta^5,\beta,\frac{\xi_3}{4\xi_4},1],
\ea
\]
and using
\eqref{eq:V2_1}, \eqref{eq:V1b_1} and  \eqref{eq:V3b_1}, with \eqref{eq:V_1b}, we have

\[
\ba{rcl}
    \dot V \leq -\gamma V,\ \gamma = \mathrm{min}[\frac{c_3}{3\beta},\xi_3\beta^2,\eta\frac{\mu_n}{2}].
\ea
\]

We can derive 
$\|\xch\|$ and $\|\xpi\|$ are exponentially convergent to the origin by \eqref{eq:V_1b} and so is $\|\xl\|$  by \eqref{eq:HK}. 
With the definition $\xl(t)=\xb(t)-(\mathbf{1}_n\otimes \mathbf{I}_d)s^\ast$, 
Theorem 
\ref{thm-1}.ii) holds.



\subsection{Proof for Theorem \ref{thm-2}}
\label{pr:The2}

In \cite{XY-CCFD}, the following algorithm is proved to be equal to Flow \eqref{eq:Algorithm2_im}.
\begin{equation}
\ba{rcl}\label{eq:Algorithm2}
\dot{\sigma}_i&=&\mathcal{C}(x_i-\sigma_i,t)\\
\dot{x}_{i}&=&-\alpha\sum^{n}_{j=1}L_{ij} {x}_{j,c}-\beta v_{i}-\eta \nabla f_i(x_i) \\
\dot{v}_{i}&=&{\beta\sum^{n}_{j=1}L_{ij}{x}_{j,c}}\\
{x}_{i,c}&=& \sigma_i + \mathcal{C}(x_i-\sigma_i,t),
\ea
\end{equation}
where the initial condition is $\sum_{i=1}^n v_i(0)=\mathbf{0}_d$ and $\sigma_i(0)=\mathbf{0}_d,\forall 
i\in\mathrm 
V$. 

Next, we will prove the validity of \eqref{eq:Algorithm2} instead.
 Flow \eqref{eq:Algorithm2} can be written as 
\begin{equation}
\ba{rcl}\label{eq:Algorithm2_tight}
\dot{\sigmab}&=&\Cb(\xb-\sigmab,t) \\
\dot{\xb}&=&-[\alpha \Lb{\xb_c}+\beta \vb+\eta \Fb(\xb)] \\
\dot{\vb}&=&{\beta \Lb{\xb_c}}\\
\xb_c&=&\sigmab+\Cb(\xb-\sigmab,t).
\ea
\end{equation}
where $\sigmab(t)=[\sigma_1^T(t),...\sigma_n^T(t)]$ and $\Cb(\xb-\sigmab):=[\mathcal{C}^T(x_1-\sigma_1)...\mathcal{C}^T(x_n-\sigma_n)]^T.$


As $f(x)$ is convex, there exists some $s^\ast\in\mathbb{R}^d$ that $\nabla f(s^\ast)=0$. Then define the state error $\tilde{\xb}(t):=\xb(t)-\xb^\ast$,  $\tilde{\sigmab}(t):=\sigmab(t)-\sigmab^\ast$, 
$\tilde{\vb}(t):=\vb(t)+\frac{\eta \Fb(\Hb\xb(t))}{\beta}$. 
Taking the time derivative of the state errors along \eqref{eq:Algorithm2_tight} yields  
\begin{equation}
\ba{rcl}\label{eq:Algorithm2_tight_0}
\dot{\sil}&=&\Cb(\xl-\sil,t) \\
\dot{\xl}&=&-[\alpha \Lb\xhl+\beta \vl+\eta [\Fb(\xb)-\Fb(H\xb)]] \\
\dot{\vl}&=&\beta \Lb{\xhl}+\frac{\eta}{\beta}\dot \Fb(H\xb)\\
\xhl&=&\sil+\Cb(\xl-\sil,t).
\ea
\end{equation}


By \eqref{eq:HL}, \eqref{eq:HK} and \eqref{eq:KL}, system \eqref{eq:Algorithm2_tight_0} becomes
\begin{equation}
\ba{rcl}\label{eq:Algorithm2_tight_0_f}
\dot{\sil}&=&\Cb(\xl-\sil,t)\\
\begin{bmatrix}
    \mathbf{\dot{\tilde{x}}}_\perp\\
    \dot{\vl}
\end{bmatrix}&=&\Nb
\begin{bmatrix}
    \mathbf{\xhl}\\
    {\vl}
\end{bmatrix}+\Mb(\xb)
\\
\mathbf{\dot{\tilde{x}}}_\parallel&=&- \Hb\eta \Fb(\xb) \\
\xhl&=&\sil+\Cb(\xl-\sil,t),
\ea
\end{equation}

We choose $V_1(\xch,\vl)=\|\begin{array}{c}
    \xch\\
    \vl
\end{array}\|^2_\Qb$,
where we define $\Qb:=\frac{1}{2}\begin{bmatrix}
    \mathbf{I}_{nd} & \Pb\\
    \Pb & \frac{\alpha+\beta}{\beta}\Pb
\end{bmatrix}$, It can be easily concluded that 
\begin{equation}
\ba{c}
\Nb^T\Qb+\Qb\Nb=-\beta\begin{bmatrix}
    -\alpha \Lb+\beta \Kb & \beta \Kb \\
    -\beta\mathbf{I}_{nd} & -\beta \Pb
\end{bmatrix}.
\ea
\end{equation}
Then, computing the time-derivative of $V_1$ along \eqref{eq:Algorithm2_tight_0_f} yields

\begin{equation}
\label{eq:V1a_2}
\ba{rcl}
\dot{V}_1 
&\leq& -\frac{1}{2}\alpha \lambda_2\|\xch\|^2 +  (\frac{1}{2}\lambda_n\alpha +\lambda_n\beta)\|\xhl-\xl\|^2\\&&
-(\frac{3}{4}\beta- 4\frac{\eta}{\lambda_2}-\frac{\eta\beta}{4})\|\vl\|^2_{\Pb}\\
&&+\beta\|\xhl\|^2_\Kb + (\frac{\beta}{4}+{\eta}+\frac{5\eta L_f^2}{8})\|\xch\|^2\\
&&+(\frac{\eta^3 L_f^2}{2\beta^2\lambda_2^2}+\frac{1}{8}\eta+\frac{\eta^3L_f^2(\alpha+\beta)^2}{\lambda_2\beta^5})(\|\Hb\Fb(\xb)\|^2),

\ea
\end{equation}
which is obtained by \eqref{eq:HK}, \eqref{eq:KL}, 
\eqref{eq:G_xHx} ,\eqref{eq:PI}, \eqref{eq:PL}, the fact $
\lambda_2 \Kb\leq \Lb\leq\lambda_n \Kb
$ and $\Kb\xch=\xch$.

As $\dot{x}_e=-\mathcal{C}(x_e,t)$ is exponentially convergent at the zero equilibrium, where $x_e\in\mathbb{R}^d$, then there exists a Lyapunov function $V_e(x_e,t):\mathbb{R}^{d}\times\mathbb{R}_+\rightarrow\mathbb{R}$ which satisfies 
\begin{equation}
\ba{l}\label{eq:Ve_2}
c_1\|x_e\|^2 \leq V_{e}(x_e,t)\leq c_2\|x_e\|^2\\
\frac{\partial V_{e}}{\partial t} - \frac{\partial V_{e}}{\partial x_e} \mathcal{C}(x_e,t)  \leq -c_3\|x_e\|^2\\
\|\frac{\partial V_{e}}{\partial x_e}\| \leq c_4 \|x_e\|
\ea
\end{equation}
for some $c_1,c_2,c_3,c_4>0$.

We choose $V_2(\xl-\sil,t):=V_2(\xb-\sigmab,t)=\sum_{i=1}^n V_e(x_i-\sigma_i,t)$, then 
\begin{equation}
\ba{rcl}\label{eq:V2_2}
\dot{V}_2 &=& \frac{\partial V_{2}}{\partial t} + \frac{\partial V_{2}}{\partial (\xl-\sil)} [\dot{\xl}-\Cb(\xl-\sil,t)]\\ 
&\leq& -c_3 \|\xl-\sil\|^2 + c_4\|\xl-\sil\|\\
&&\|\alpha \Lb\xhl+\beta \vl+\eta[\Fb(\xb)-\Fb(\Hb\xb)]\| \\
&\leq& -(c_3-c_4\alpha/r-
c_4\beta/r-c_4\eta/r) \|\xl-\sil\|^2 \\
&&+ c_4\alpha r\lambda_n^2\|\xhl\|_\Kb^2+c_4\beta r\lambda_n\|\vl\|_\Pb^2+c_4\eta rL_f^2\|\xch\|^2,
\ea
\end{equation}
where the first inequality is obtained by \eqref{eq:Ve_2} and the last inequality is obtained by \eqref{eq:PI}, \eqref{eq:G_xHx} and Young's Inequality, where $r>0$ is a parameter which will be determined later.

Before we proceed to the proof of statements i) and ii),  in the following we introduce some useful parameters, independent of $\alpha$, $\beta$, $r$ and $\eta$.
\[
\ba{rcl}
&&\xi_0=\frac{8L_f^4}{\lambda_2\mu_n}+\frac{L_f^4}{\lambda_2^2\mu_n}+\frac{L_f^2}{4\mu_n},\quad\xi_1=\frac{1}{2} \lambda_2,\quad \\

&&\xi_2=1+\frac{9L^2_f}{8},\quad\xi_3=2c_4\lambda_n^2,\quad\\
&&\xi_4=\frac{9}{4},\quad\xi_5=\frac{1}{2},\quad\xi_6=\frac{4}{\lambda_2},\quad\xi_7=c_4\lambda_n,\\
&&\xi_8=3c_4,\quad\xi_9=\theta(\frac{3}{2}\lambda_n+\frac{8L_f^4}{\lambda_2}+2+2c_4\lambda_n^2),\\
&&\xi_{10}=\frac{4L_f^4}{\lambda_2}+\frac{L_f^4}{2\lambda_2^2},\quad\xi_2'=\xi_2+\frac{4L_f^4}{\lambda_2}+\frac{L_f^4}{2\lambda_2^2}.
\ea
\]

\emph{Proof of i).}
Define $V_{3,a}=f(\frac{1}{n}((\mathbf{1}_n^T\otimes\mathbf{I}_d)\xl))-f^\ast$. To this end, we choose  $V(\xl,\vl)= V_1(\Kb\xl,\vl)+V_2(\xl-\sil,t)+V_{3,a}(\xl)$. With \eqref{eq:Ve_2}, it is easy to prove that
$V$ satisfies 
\begin{equation}
\ba{rcl}\label{eq:V_2a}
V\geq\frac{1}{4}\|\xch\|^2+\frac{1}{2}\|\vl\|^2_\Pb+ c_1\|\xl-\sil\|^2.

\ea
\end{equation}
as we choose $\alpha\geq2\beta$. 

First, let's introduce some facts 
\begin{equation}
\ba{rcl}\label{eq:fact1}
\|\xhl-\xl\|^2=\|\xl-\sil-\Cb(\xl-\sil,t)\|^2\leq \theta\|\xl-\sil\|^2
\ea
\end{equation}
\begin{equation}
\ba{rcl}\label{eq:fact2}
\|\xhl\|^2_\Kb\leq2\|\xl-\xhl\|^2 + 2\|\xch\|^2\leq2\theta\|\xl-\sil\|^2 + 2\|\xch\|^2.
\ea
\end{equation}
for $\theta:=2+2L_c^2>0$ because $\mathcal{C}(x_e,t)$ is a ST compressor. 

By fixing
\[
\ba{rcl}
 &&r=\mathrm{min}[\frac{\xi_1}{2\xi_3},\frac{\xi_5}{2\xi_7},1], \ \alpha\leq \mathrm{min}[\frac{c_3r}{3\xi_8}, \frac{c_3}{3\xi_9}],\\
 &&\beta\leq \mathrm{min}[\frac{\alpha}{2},\frac{\xi_1\alpha}{4\xi_4}], \  \eta\leq \mathrm{min}[\beta^5,\beta,\alpha^{-2},1,\frac{\xi_1}{8\xi_2},\frac{\xi_5}{4\xi_6},
{\frac{3}{8\xi_{10}}}],
 \ea
\]
and using \eqref{eq:V3a_1}, \eqref{eq:V1a_2} and \eqref{eq:V2_2},
\[
    \ba{rcl}
    \dot V&\leq& -\frac{\xi_1 \alpha}{8}\|\xch\|^2-\frac{\xi_5\beta}{4}\|\vl\|_\Pb^2\\
    &&-\frac{c_3}{3}\|\xl-\sil\|^2.
    \ea
\]

Similar to the proof of \ref{thm-1}.i). The prove of Theorem 
\ref{thm-2}.i) is complete.

\emph{Proof of ii).}
As $f(x)$ is strongly convex with $\mu>0$, \eqref{eq:G_mu} holds. 

By \eqref{eq:G_dHx} and \eqref{eq:V1a_2},
\begin{equation}
\label{eq:V1b_2}
\ba{rcl}
\dot{V}_1  
&\leq& -\frac{1}{2}\alpha \lambda_2\|\xch\|^2 +  (\frac{1}{2}\lambda_n\alpha +\lambda_n\beta)\|\xhl-\xl\|^2\\&&
-(\frac{3}{4}\beta- 4\frac{\eta}{\lambda_2}-\frac{\eta\beta}{4})\|\vl\|^2_{\Pb}\\
&&+\beta\|\xhl\|^2_K + (\frac{\beta}{4}+{\eta}+\frac{5\eta L_f^2}{8})\|\xch\|^2\\
&&+(\frac{\eta^3 L_f^4}{2\beta^2\lambda_2^2}+\frac{L_f^2}{8}\eta+\frac{\eta^3L_f^4(\alpha+\beta)^2}{\lambda_2\beta^5})(\|\xch\|^2+\|\xpi\|^2)
\ea
\end{equation}

Define $V_{3,b}(\xpi):=\frac{1}{2}\|\xpi\|^2$ and  define the Lyapunov functions of system \eqref{eq:Algorithm2_tight_0_f} $V(\xl,\vl)= V_1(\Kb\xl,\vl)+V_2(\xl-\sil,t)+2\xi_0V_{3,b}(\Hb\xl)$. 
As we choose $\alpha\geq2\beta$, it is easy to prove that $V$ is positive definite. In fact,
\begin{equation}
\ba{rcl}\label{eq:V_2b}
V\geq\frac{1}{4}\|\xch\|^2+\frac{1}{2}\|\vl\|^2_\Pb+ c_1\|\xl-\sil\|^2 + \xi_0\|\xpi\|^2.
\ea
\end{equation}



By fixing
\[
\ba{rcl}
&&r=\mathrm{min}[\frac{\xi_1}{2\xi_3},\frac{\xi_5}{2\xi_7},1], \ \alpha\leq \mathrm{min}[\frac{c_3r}{3\xi_8}, \frac{c_3}{3\xi_9}], \\&&\beta\leq \mathrm{min}[\frac{\alpha}{2},\frac{\xi_1\alpha}{4\xi_4}], \ \eta\leq \mathrm{min}[\beta^5,\beta,\alpha^{-2},1,\frac{\xi_1}{8\xi_2'},\frac{\xi_5}{4\xi_6}],
\ea
\]
and using
\eqref{eq:V2_2}, \eqref{eq:V1b_2},  \eqref{eq:V3b_1}, \eqref{eq:fact1} and \eqref{eq:fact2}, 
with \eqref{eq:V_2b}, we have

\[
\ba{rcl}
    \dot V \leq -\gamma V,\ \gamma = \mathrm{min}[\frac{\xi_1 \alpha}{2},\frac{\xi_5\beta}{2},\frac{c_{3}}{3c_1},\eta\frac{\mu_n}{2}].
\ea
\]

Similar to the proof of Theorem 
\ref{thm-1}.ii),
Theorem 
\ref{thm-2}.ii) holds.


\end{document}